\documentclass[11pt]{article}
\usepackage{deluxetable,amsmath,fullpage,graphicx,booktabs,palatino,epsfig,fancyhdr,color,amssymb,relsize,setspace}

\renewcommand{\headrule}{{\hrule width\headwidth height\noheadrule}}

\newcommand{\appsection}[1]{\let\oldthesection\thesection
\renewcommand{\thesection}{Appendix \oldthesection}
\section{#1}\let\thesection\oldthesection}

\begin{document}

\begin{figure*}[h]
\begin{center}
\vspace{-1.0cm}
\includegraphics[height=4.0cm]{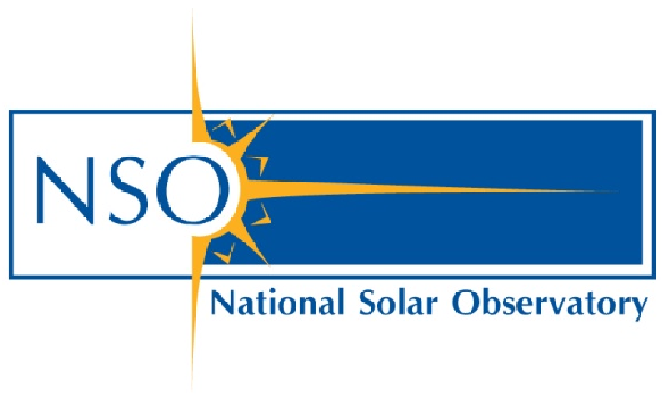}
\end{center}
\end{figure*}

\begin{center}

\vspace{2.0cm}

\begin{LARGE}
\textbf{Camera Gap Removal in SOLIS/VSM Images}
\end{LARGE}

\vspace{3.5cm}

\begin{large}

Andrew R. Marble, Lorraine Callahan \& Alexei A. Pevtsov

\vspace{0.5cm}

National Solar Observatory

\vspace{3.5cm}

October 14, 2013

\vspace{4.5cm}

\hrule

\vspace{1.0cm}

Technical Report No. \textbf{NSO/NISP-2013-003}

\end{large}

\end{center}

\clearpage
\renewcommand{\headrule}{{\hrule width\headwidth 
height\headrulewidth\vskip0.5cm}}
\fancyhead{}
\fancyhead[L]{Camera Gap Removal in SOLIS/VSM Images}
\fancyhead[R]{\thepage}
\fancyfoot{}
\thispagestyle{fancy}

\vspace*{-0.40cm}

\begin{center}
\begin{Large}
\bf{Abstract}
\end{Large}
\end{center}

\begin{quote}
The Vector Spectromagnetograph (VSM) instrument on the Synoptic Optical 
Long-term Investigations of the Sun (SOLIS) telescope is capable of 
obtaining spectropolarimetry for the full Sun (or a select latitudinal range)
with $1''$ spatial 
resolution and 0.05\AA\ spectral resolution.  This is achieved by scanning the Sun
in declination and building up spectral cubes for multiple polarization states,
utilizing a beamsplitter and two separate $2k\times2k$ CCD cameras.
As a result, the eastern and 
western hemispheres of the Sun are separated in preliminary VSM images by a 
vertical gap with soft edges and variable position and width.
Prior to the comprehensive analysis presented in this document,
a trial-and-error approach to removing the gap had yielded an
algorithm that was inconsistent, undocumented, and responsible
for incorrectly eliminating too many image columns.  Here we describe, in detail,
the basis for a new, streamlined, and properly calibrated prescription
for locating and removing the gap that is
correct to within approximately $1''$ (one column).
\end{quote}

\vspace*{-0.10cm}

\section{The Origin of the Gap}

\begin{figure*}[h]
\begin{center}
\includegraphics[width=\linewidth]{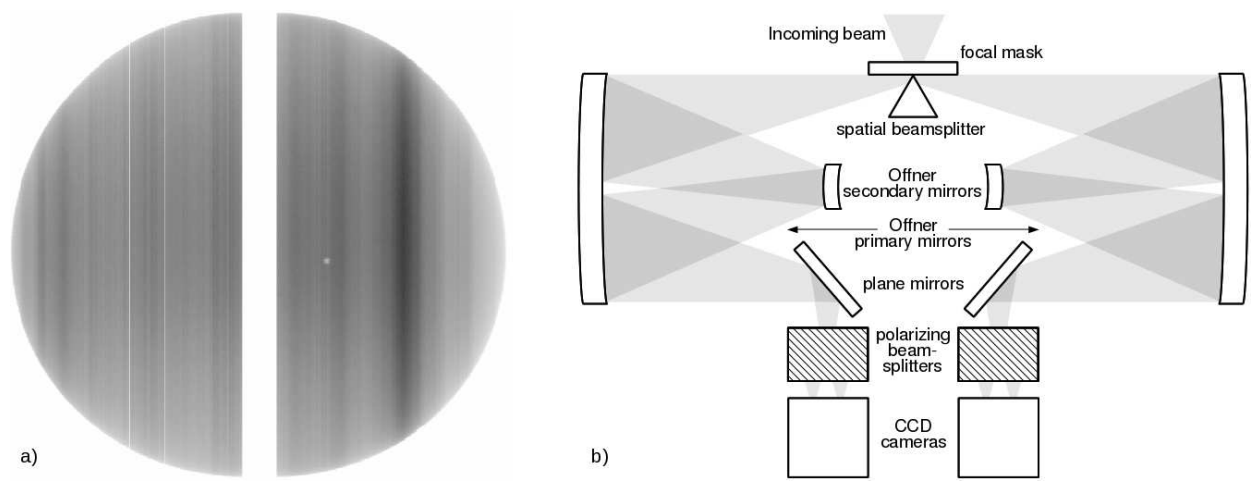}
\caption{\small-- \emph{(a) Level-1 6302L intensity image (without flat-field corrections) exhibiting the gap.
\,\,\,(b) Schematic of the reimaging system within the VSM on SOLIS.}}
\label{fig_gap}
\end{center}
\end{figure*}

Initial images made with the Vector Spectromagnetograph (VSM) instrument 
on the Synoptic Optical Long-term Investigations of the Sun (SOLIS) telescope
include a gap between the eastern and western hemispheres of the 
Sun (see Figure~\ref{fig_gap}a).  This gap corresponds to a physical separation
in the focal plane of two halves of the solar disk image, where the width
is determined by the positions of the cameras and optical train elements
beyond the beamsplitter.  As illustrated in Figure~\ref{fig_gap}b, the (already 
dispersed) beam is spatially split into halves 
that are imaged onto two CCD cameras.  To ensure no image 
loss, the cameras are illuminated such that there is a buffer on the order 
of tens of pixels between the spatially split edge of the beams and the 
corresponding end of the cameras.  Thus, a gap devoid of illumination is 
present in the center of each row (Figure~\ref{fig_dailychanges}a) 
constructed by juxtaposing
the camera images for each scan-line during Level 0-1 processing.

\newpage

\vspace*{-0.5cm}

\section{Relative Changes in the Gap Position/Size}\label{sec_instability}

\begin{figure*}[h]
\includegraphics[width=6.5in]{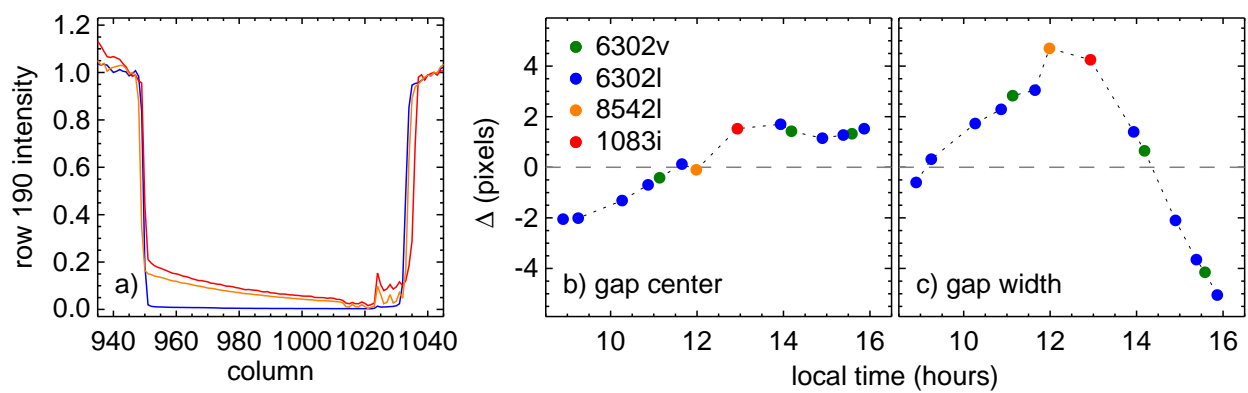}
\caption{\small-- \emph{Examples of the gap profile (a) from three flats taken on February 23, 2011
(separately normalized on both the left and right sides of the gap)
  and changes in the gap center (b) and width (c) from additional flats taken 
 throughout the same day.}}
\label{fig_dailychanges}
\end{figure*}

Cursory inspection of the gap profile seen in flats shows that
the position and width of the gap are not constant, but rather change 
both from day-to-day and throughout the day 
(due to various effects that likely include flexure and
thermal expansion in the optical train beyond the beamsplitter; see 
Figure~\ref{fig_gap}b).  As an example, these two properties of
the gap were measured (using a fiducial definition of the gap as 
where the normalized intensity falls below 50\%)
and plotted in Figure~\ref{fig_dailychanges}
for a day with an unusually large number of flats.  Not only is the position of
the gap shown to shift by four pixels throughout that particular day, but the width 
varied by more than twice as much during the afternoon hours (approximately three pixels an hour).

\hspace{0.5cm}

Thus, removal of the gap must be tailored for each
observation in order to properly preserve the full solar disk as well as the 
spatial relationship between the halves to the left and right of the gap.
The relatively rapid changes with time mean that identification of the gap
should be based on an observation directly rather than its 
corresponding flat (as was done in the past; see Appendix A).  More than 50\% 
of the observations prior to September 2011 were paired with a flat taken at
least 30 minutes before the mid-point of the observation, while 10\% 
were separated in time by at least an hour.

Using the \emph{central} row of the intensity image to characterize the gap further 
mitigates the significant time elapsed between the start and end of an observation.
Based on Figure~\ref{fig_dailychanges}, the gap center/width shouldn't change
by more than a half/whole pixel during the 20 minute interval between 
observing the disk center and either pole for 8542L (the longest observation).
There have been undocumented references to gaps with tilts somewhat larger than this;
however, casual inspection of the gap profile can be misleading due to lack of
reliable flat-fielding within the gap (where the cameras are not illuminated well or 
at all) and careful analysis is hindered by the 
fact that the Level 0-1 pipeline previously replaced pixels \emph{thought} to be 
well within the gap with zeroes.
In a few instances where this study uncovered gaps seemingly tilted on one side by as 
much as a pixel, it was found that the width was changing (rather than the gap center), 
due in turn to changes in the slope of the gap profile edges.  Scattered light 
as a function of the Sun's orientation with respect to the slit may have been a 
contributing factor.

\newpage

\section{Absolute Spatial Extent Measurements of the Gap}\label{sec_measurements}

Due to the finite sharpness of the spatial beamsplitter tip, the transition
between fully-illuminated and unilluminated portions of the cameras spans 
several pixels.  Additionally, the presence of scattered light
degrades, to varying degrees, the slope of the gap profile edges.
In order to distinguish columns containing diminished, but valid, data
from those inside the gap that are masked by stray light,
the true spatial extent of the gap was carefully and independently measured for ten 
6302\AA, 8542\AA, and 10830\AA\ full-disk observations
(see Table~\ref{tab_widths}).  This was done
by comparing Level-1 data containing the gap to
nearly simultaneous observations of the Sun taken at similar wavelengths 
with other instruments employing a single camera.  
HMI 6173\AA\ magnetograms 
(hmi.M\_45s.\emph{yyyymmdd}\_\emph{hhmmss}\_TAI.2.magnetogram.fits)
and ISOON 10830\AA\ (line center minus wing) intensity images
(\emph{yyyymmddhhmmss}yc.fits) were used as references for the 6302L/8542L 
(the wing of the 8542\AA\ line is essentially photospheric) magnetograms 
and 1083I equivalent width images, respectively.
Both the HMI and ISOON pixel scales (0.5'' and 1.1'') are
comparable to the VSM's 1'' pixels, and all of the reference images 
were made within three minutes of the midpoint of the corresponding 
VSM observations.

For each pair of VSM and reference images, the latter was rotated, resampled, 
and shifted in order to achieve rough alignment.  
Then, vertical $200\times150$ pixel regions to the left and right of the disk 
center (and the gap) were used to cross-correlate the two images.  
The difference in the cross-correlation peak coordinates between the left and 
right regions is equal to the number of additional columns present in the VSM
image due to the gap.  The gap widths (provided in 
Table~\ref{tab_widths} and displayed in Figure~\ref{gap_profiles}) 
were also measured using two other methods: centroiding of discrete features 
and visual inspection.  In addition to being much more time-intensive, these
approaches utilize less spatial information and are more sensitive to systematic
biases such as geometry distortions across the disk.  However, they
provided a useful consistency check.

\begin{deluxetable}{c c c c c c c c c c c}
\tabletypesize{\scriptsize}
\tablecolumns{11}
\tablewidth{0pc}
\tablecaption{Measured Gap Widths.\label{tab_widths}}
\tablehead{
\multicolumn{2}{c}{} &
\multicolumn{2}{c}{Time} &
\colhead{} &
\colhead{$|$} &
\multicolumn{5}{c}{\emph{continued...}} \\
\cline{3-4}\\[-0.25cm]
\colhead{Type} &
\colhead{Date} &
\colhead{VSM} &
\colhead{Reference} &
\colhead{Gap} &
\colhead{$|$} & 
\multicolumn{5}{c}{} \\
\colhead{} &
\colhead{(yyyy.mm.dd)} &
\colhead{(hh:mm)} &
\colhead{(hh:mm:ss)} &
\colhead{(pixels)} &
\colhead{$|$} &
\colhead{} &
\colhead{(yyyy.mm.dd)} &
\colhead{(hh:mm)} &
\colhead{(hh:mm:ss)} &
\colhead{(pixels)}
}
\startdata
 6302L & 2011.05.06 & 14:50 & 14:54:45 & 76.4 & $|$ & 8542L & 2011.05.19 & 15:15 & 15:36:45 & 83.7\\
 6302L & 2011.04.27 & 14:34 & 14:39:00 & 76.2 & $|$ & 8542L & 2010.12.06 & 15:53 & 16:15:00 & 77.2\\
 6302L & 2011.04.11 & 15:13 & 15:18:00 & 79.1 & $|$ & 8542L & 2010.08.27 & 15:23 & 15:45:00 & 80.3\\
 6302L & 2011.04.05 & 14:42 & 14:47:15 & 73.8 & $|$ & 8542L & 2010.08.02 & 21:24 & 21:45:45 & 77.6\\
 6302L & 2011.04.01 & 14:42 & 14:47:15 & 75.0 & $|$ & 8542L & 2010.07.04 & 22:20 & 22:42:00 & 74.4\\
 6302L & 2011.03.28 & 14:50 & 14:55:30 & 76.2 & $|$ & 1083I & 2011.05.28 & 18:02 & 18:09:22 & 80.6\\
 6302L & 2011.03.12 & 16:34 & 16:39:00 & 79.6 & $|$ & 1083I & 2011.05.22 & 15:35 & 15:39:22 & 81.5\\
 6302L & 2011.03.02 & 15:47 & 15:53:15 & 77.0 & $|$ & 1083I & 2011.05.14 & 18:04 & 18:09:22 & 82.6\\
 6302L & 2010.12.08 & 15:41 & 15:45:45 & 73.9 & $|$ & 1083I & 2011.04.01 & 16:42 & 16:49:22 & 79.9\\
 6302L & 2010.08.09 & 15:57 & 16:02:15 & 78.4 & $|$ & 1083I & 2011.03.28 & 17:35 & 17:39:22 & 85.1\\
 8542L & 2011.11.10 & 18:10 & 18:32:15 & 83.3 & $|$ & 1083I & 2011.03.14 & 16:46 & 16:49:23 & 83.0\\
 8542L & 2011.10.18 & 15:28 & 15:50:15 & 79.2 & $|$ & 1083I & 2011.03.05 & 15:41 & 15:49:22 & 78.3\\
 8542L & 2011.10.09 & 16:24 & 16:45:45 & 83.5 & $|$ & 1083I & 2011.01.27 & 19:51 & 19:59:22 & 82.7\\
 8542L & 2011.10.03 & 15:25 & 15:47:15 & 80.6 & $|$ & 1083I & 2011.01.24 & 20:54 & 20:59:22 & 80.4\\
 8542L & 2011.06.23 & 15:05 & 15:27:00 & 79.7 & $|$ & 1083I & 2011.01.19 & 18:23 & 18:29:22 & 82.9\\
\enddata
\end{deluxetable}

\newpage

\begin{figure*}[h]
\centerline{
  \mbox{\includegraphics[width=\linewidth]{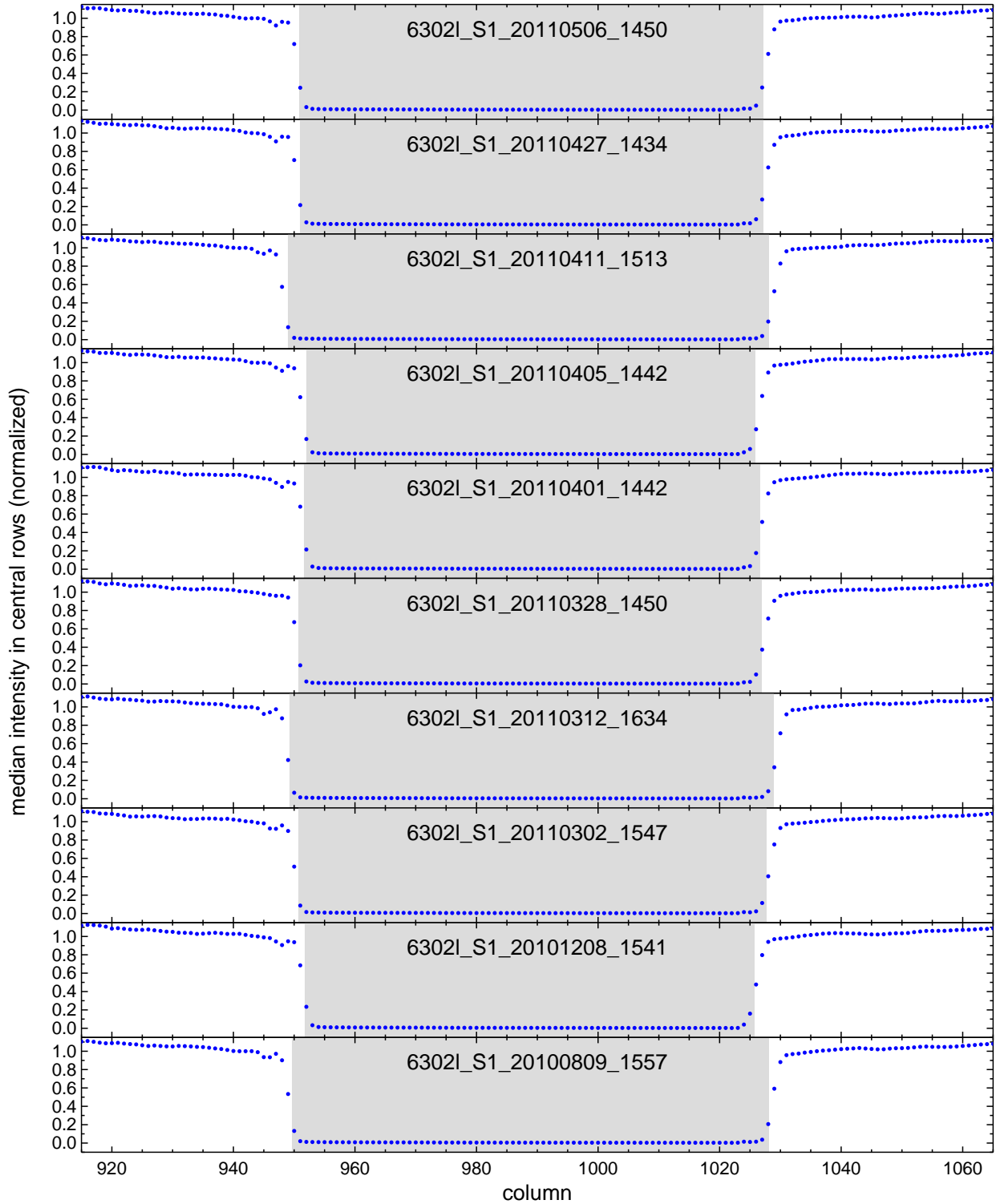}}
}
\caption{\small-- \emph{Taken from the un-flat-fielded intensity images, 
these gap profiles are the median of the 20 most central 
rows normalized (separately to the left and right of column 990) 
by the median of the intensities within 3-13 pixels of the gap.
The independently measured gap widths listed in Table~\ref{tab_widths}
are indicated by the shaded regions that have been centered on each gap profile.}}
\label{gap_profiles}
\end{figure*}

\setcounter{figure}{2}
\begin{figure*}
\centerline{
  \mbox{\includegraphics[width=\linewidth]{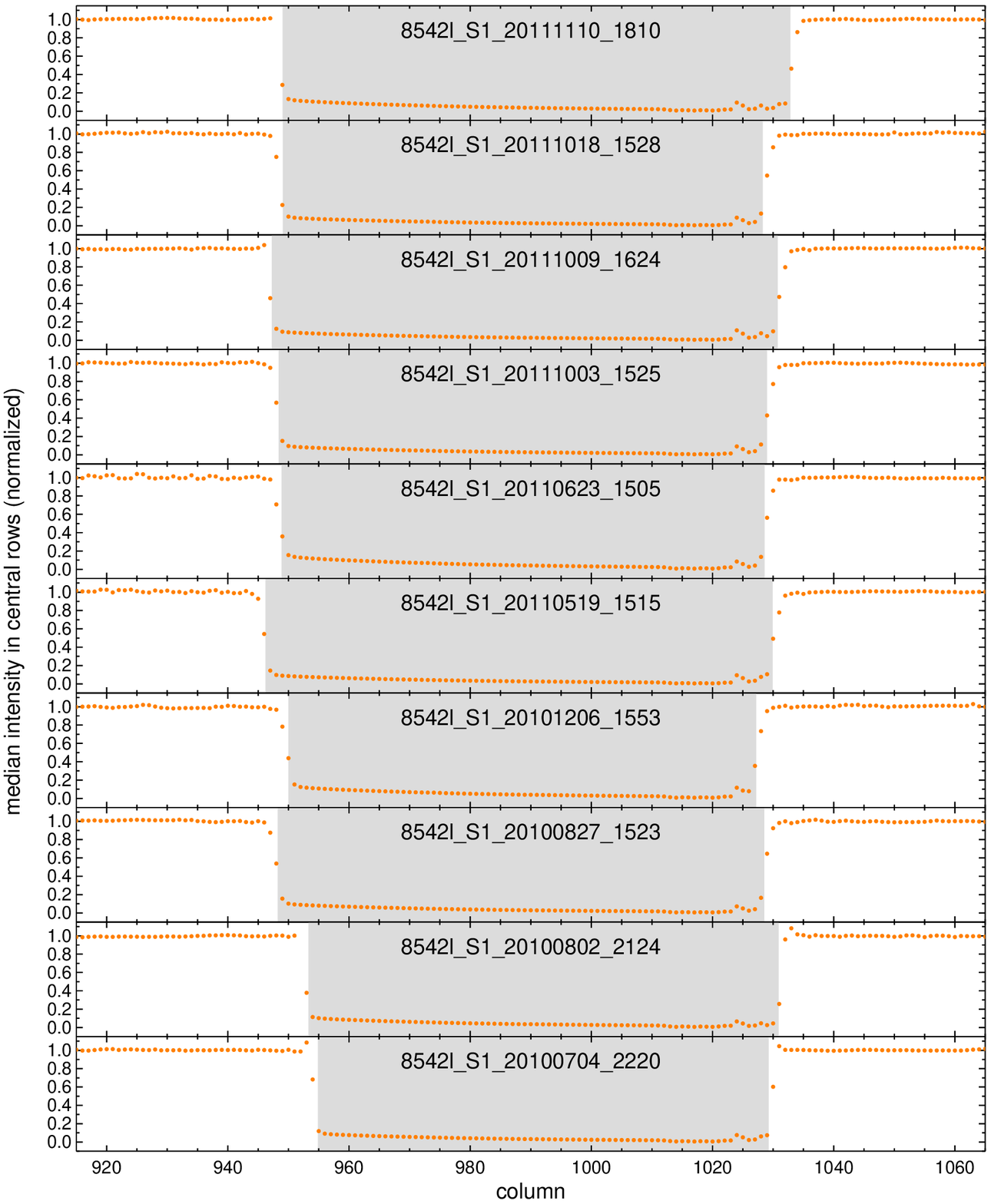}}
}
\caption{continued...}
\end{figure*}

\setcounter{figure}{2}
\begin{figure*}
\centerline{
  \mbox{\includegraphics[width=\linewidth]{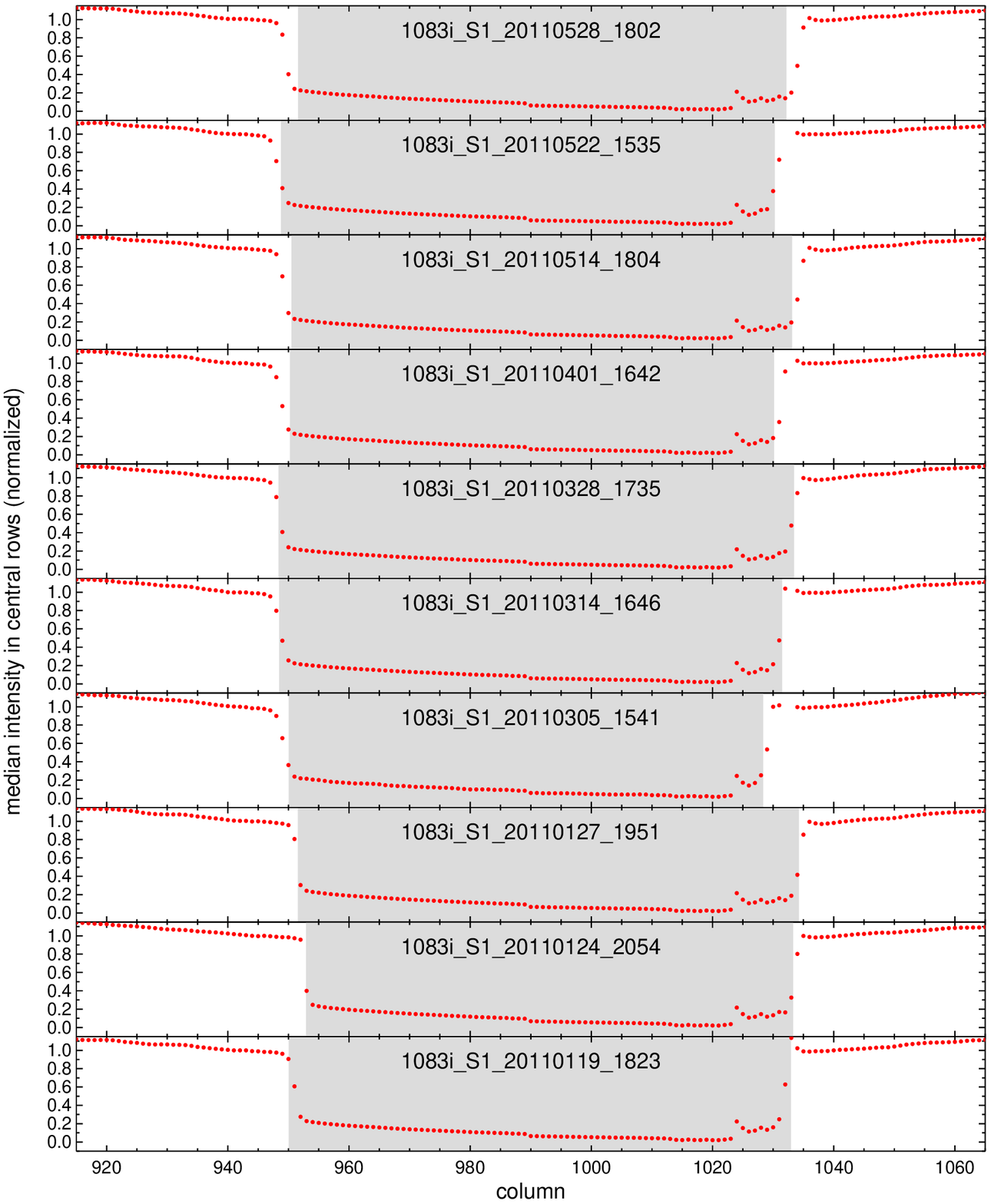}}
}
\caption{continued...}
\end{figure*}

\newpage
\clearpage
\pagebreak

\vspace*{-0.5cm}

\section{Gap Profile Intensity Thresholds}\label{sec_thresholds}

Figure~\ref{gap_profiles} shows the 
central row gap profiles for the thirty observations with independently 
measured gap widths. These profiles come from 
un-flat-fielded intensity images produced specifically for this analysis.  
Flat-fielding is not valid inside the gap and results in computationally 
distorted profiles due to division of very small numbers by other very small 
numbers.  Each undistorted gap profile is the median of the twenty most 
central rows, normalized separately to the left and the right of column 990
by the median of 10 columns just outside of the gap.  
The edges of the shaded regions corresponding to the gap width measurements 
in Table~\ref{tab_widths} similarly intersect the profiles 
near 50\% of the fully-illuminated
intensity (there is greater variance for the 1083I images due to less certain
gap width measurements related to decreased contrast and worse seeing
in both the infrared VSM observations and the corresponding ground-based reference
images).

\begin{figure*}[hb]
\centerline{
  \mbox{\includegraphics[width=\linewidth]{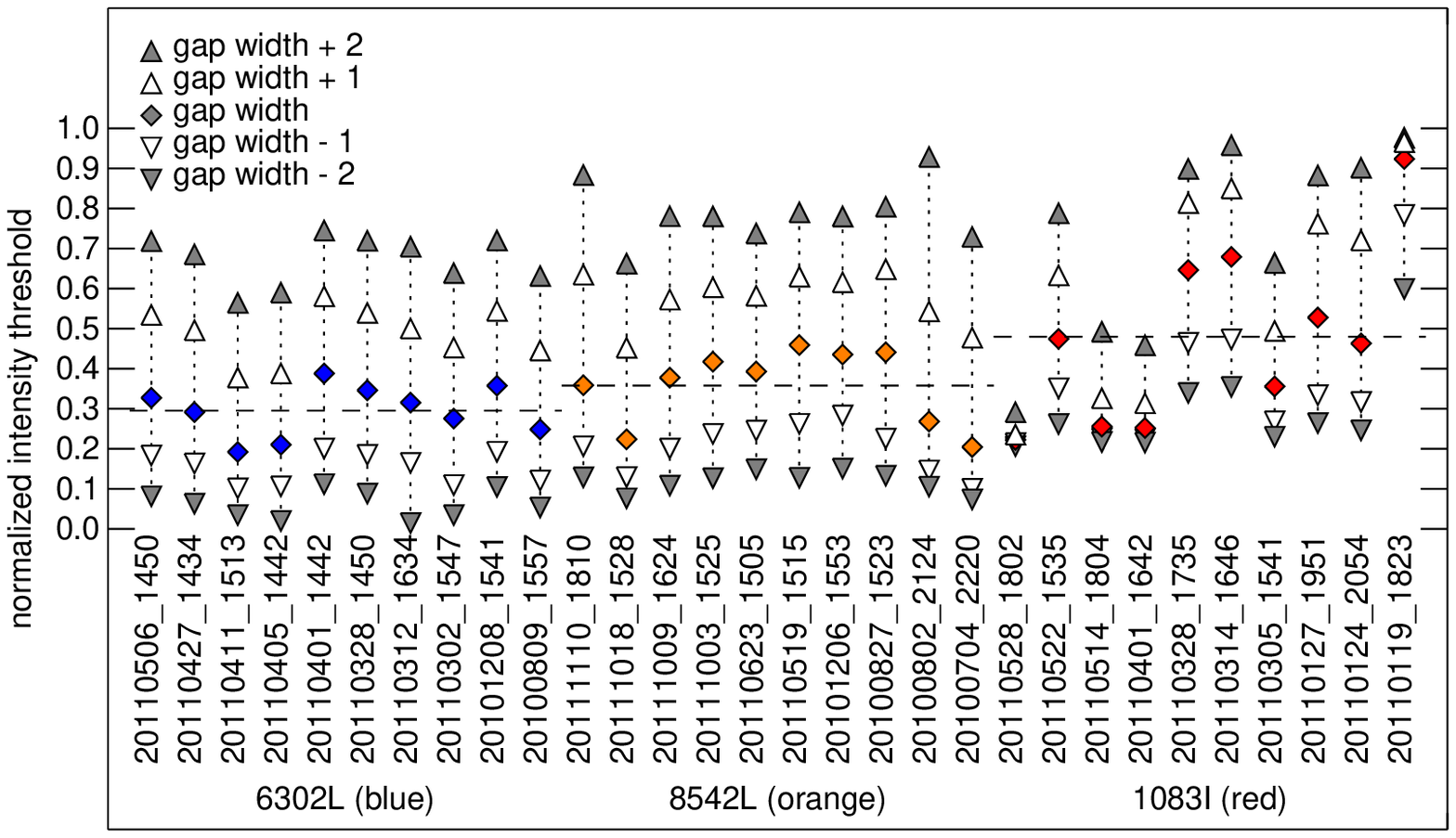}}
}
\caption{\small-- \emph{Normalized intensity thresholds corresponding to the gap widths 
independently measured for ten 6302L (blue), 8542L (orange), and 1083I (red) observations.}}
\label{gap_threshold}
\end{figure*}

The actual (interpolated) intensity thresholds corresponding to the measured 
gap widths are plotted as diamonds in Figure~\ref{gap_threshold}.  Similarly, 
values yielding widths too large and too small by one or two columns are 
included (the latter become indistinguishable once the flat parts of the 
profile are reached).  There is excellent agreement between the 6302L 
and 8542L observations, where the mean thresholds of 30\% and 36\% match the measured gap
widths to the nearest integer in all cases.  The poorer agreement between the 
10830I observations is consistent with the comparatively less certain gap
width measurements and is mitigated by once again adopting the mean threshold
(48\%).  The correlation between threshold value and wavelength is expected due to 
relative contributions from scattered light.

\newpage

\section{A New Gap Removal Prescription}

Beginning with the Level-0 to Level-1 pipeline versions (indicated by PROVER0) 
11.1014, 12.0403, and 13.1001, respectively, for 6302L/1083I, 8542L, and 6302V 
data, the 
{\tt GAPCOL1} and {\tt GAPCOL2} keywords in the Level-1 headers correctly 
identify the first and last extraneous columns inside the gap.  These values 
are determined using the following prescription, which, based on 
\S~\ref{sec_measurements} and \S~\ref{sec_thresholds} yields gap widths that 
are correct to within approximately $1''$ (one column).
\begin{quote}
\begin{small}
\onehalfspacing

Take the median profile
for the 20 most central rows in the un-flat-fielded intensity image,
and assume a threshold $X$ of 0.30 for 6302[L/V], 0.36 for 8542L, or 
0.46 for 1083I (see \S~\ref{sec_thresholds}).
Start at the index of the central column plus 20, and move towards
lower column numbers.  {\tt GAPCOL1} is then the first (one-indexed) column
$i$ to satisfy the condition that the intensities in columns $i-2$
and $i-1$ are greater than $X$ times the median of the intensities in
columns $i-13:i-3$ whereas columns $i$ and $i+1$ have intensities
less than the same threshold.  Symmetrically, {\tt GAPCOL2} is the first
(one-indexed) column $j$ that is greater than the central column index
minus 20 where
columns $j-1$ and $j$ have intensities less than $X$ times the median
of the intensities in columns $j+3:j+13$ while columns $j+1$ and $j+2$
are greater than the same threshold.  
\end{small}
\end{quote}
\vspace{-0.5cm}
\singlespacing 
If the input Level-1 file predates these PROVER0 values, then
{\tt GAPCOL1} and {\tt GAPCOL2} are redefined during Level-1 to Level-2
processing by applying the algorithm above to a 
single central row of the flat-fielded intensity image (essentially
the same prescription described in \ref{app_oldmethod}, but utilizing the
newly determined relative intensity thresholds).

\begin{figure*}[hb]
\centerline{
  \mbox{\includegraphics[width=5.9in]{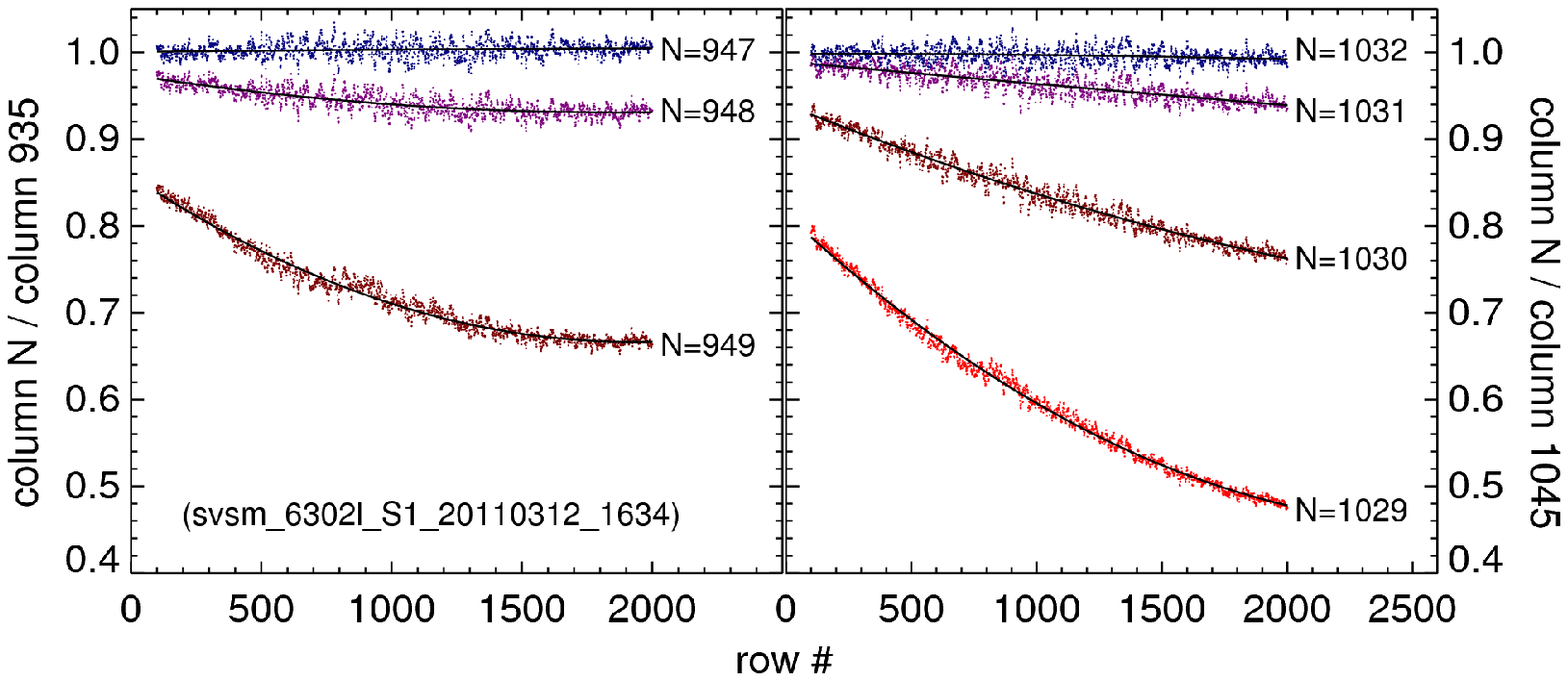}}
}
\caption{\small-- \emph{Diminished relative intensity near the gap (the gap 
is columns $950-1028$ in this case) is corrected
for via division of the affected columns by the corresponding quadratic 
polynomial fits (black lines).}}
\label{gap_wings}
\end{figure*}

The Level-1 to Level-2 pipeline
then simply removes the extraneous columns from all fits frames and
records this fact in the header.  Columns {\tt GAPCOL1}$-5$ through 
{\tt GAPCOL1}$+4$ (the five zero-indexed columns on either side of 
where the gap was) are re-scaled in order to account for 
diminished intensity near the gap.  This diminution is not
necessarily constant for all rows (see Figure~\ref{gap_wings})
due to the changes in the gap profile wings discussed in
\S~\ref{sec_instability}.  Although scattered light (which should be 
subtracted rather than scaled) may be partially responsible for these 
changes, it is not necessarily the primary cause.  Therefore, 
each column is simply divided by
a quadratic polynomial fit to the intensity ratio of that column
divided by an unaffected column (actually the median of several 
columns) further offset from the gap.

\newpage

\newpage

\appendix

\vspace*{-0.5cm}

\appsection{Previous Gap Removal Methodology}\label{app_oldmethod}

The gap removal process described in this document went into effect in 
October 2011 for 6302L and 1083I, April 2012 for 8542L, and October 2013 
for 6302V
(due to varying, and unrelated, pipeline development considerations).  For observations 
processed to Level-1 prior to these dates , the gap was identified and 
removed as described
below.  While not well documented, this less accurate and unnecessarily 
redundant methodology is likely the by-product of successive layers of 
modification during the early development of the data processing pipeline.  
The resulting gap widths and locations were generally
in error by up to several pixels and by as many as tens of pixels in extreme
cases.  This loss of data (and the corresponding spatial information necessary
for fitting and removing geometric distortions accurately) lead to the revised
methodology that is the subject of this document.

First, during Level-0 to Level-1 processing
(except for the earliest versions of the pipeline), the 
first and last pixels inside the gap were identified and written to the 
header (and database) as {\tt GAPCOL1} and {\tt GAPCOL2}, respectively, based 
on a single row from the corresponding flat.  {\tt GAPCOL1} was
defined as the last pixel to the left of the gap center that falls below a
threshold of 0.2 times the median of the pixels to the left of the gap.
Likewise, {\tt GAPCOL2} was the last pixel to the right of the gap center to 
fall below such a threshold; however, the scale factor used was 0.4 in this 
case.

Then, for 6302L and 8542L observations, {\tt GAPCOL1} and {\tt GAPCOL2} were 
used as
starting values (if not available, default starting values were adopted 
instead) for re-determining the limits of the gap during Level-1 to 
Level-2 processing using the central row of the intensity image rather 
than the flat.  The index $i$ of the last valid pixel to the left of the gap
was defined to be the instance closest to the center of the gap where pixels
$i-2$ through $i$ all lay above 0.92 times the median of pixels 
{\tt GAPCOL1}$\,-\,30$ through {\tt GAPCOL1}$\,-\,11$.
Similarly, the index $j$ of the first valid pixel to the right of the gap
was taken to be the closest occurrence to gap center of pixels $j$ through 
$j+2$ being above the same threshold with respect to the median of pixels 
{\tt GAPCOL2}$\,+\,11$ through {\tt GAPCOL2}$\,+\,30$.

In the case of 1083I observations, re-determining the gap limits was deemed
too difficult due to ``light in the gap''.  It
should be noted that the gap profile of 1083I intensity images is not, in fact,
significantly different than the observations at other wavelengths.
Although scattered light does affect the gap profile, the perceived excess
light referred to above was really an algorithmic flat-fielding distortion 
due to, ironically, a lack of illumination inside the gap.
This was not seen at other wavelengths
due to masking that was not being applied to 1083I data.
Rather than re-determine the gap limits, the last valid pixel
to the left of the gap and the first valid pixel to the right of the gap
were defined to be {\tt GAPCOL1}$\,-\,2$ and {\tt GAPCOL2}$\,+\,2$, respectively.
The subtraction/addition of a second pixel appears to be intended to compensate 
for a systematic error in, at least, the determination of {\tt GAPCOL1}.  

The gap was then 
closed by replacing those columns with indices greater than $i$ and less than $j$ 
(using the notation from above) with a single column equal to the mean of 
columns $i$ and $j$.  This, however, was not done to 6302V observations, for 
which the gap was already closed (by discarding columns {\tt GAPCOL1} + 2 
through {\tt GAPCOL2} - 2) during Level-0 to Level-1 processing for 
logistical reasons.

\end{document}